\documentclass[prl,aps,twocolumn,floats,showpacspsfig]{revtex4}
\usepackage{amssymb}

\usepackage{epsfig}

\newcommand{\be}{\begin{equation}}
\newcommand{\ee}{\end{equation}}
\newcommand{\bea}{\begin{eqnarray}}
\newcommand{\eea}{\end{eqnarray}}

\newcommand{\s}{\sigma}

\newcommand{\la}{\langle}
\newcommand{\ra}{\rangle}
\newcommand{\rd}{\mbox{d}}
\newcommand{\ri}{\mbox{i}}
\newcommand{\re}{\mbox{e}}

\begin{document}
\title{Doped Spin Liquid: Luttinger Sum Rule and Low Temperature Order}

\author{ R. M. Konik, T. M. Rice$^{*}$ and A.  M. Tsvelik}
\affiliation{ Department of  Physics, Brookhaven National Laboratory, Upton, NY 11973-5000, USA\\
$^{*}$ Institute f\"ur Theoretische Physik, ETH-H\"onggerberg, CH-8093  Z\"urich,  Switzerland}
\date{\today}

\begin{abstract}
We analyze a model of two-leg Hubbard ladders weakly coupled by interladder 
tunneling. At half filling a semimetallic state with small Fermi pockets is induced 
beyond a threshold tunneling strength. The sign changes in the single electron Green's 
function relevant for the Luttinger Sum Rule now take place at surfaces with both zeroes 
and infinities with important consequences for the interpretation of ARPES experiments. 
Residual interactions between electron and hole-like quasi-particles cause 
a transition to long range order at low temperatures. The theory can be extended to small 
doping leading to superconducting order. 

\end{abstract}

\maketitle

\sloppy

While the properties of doped spin liquids in more than one dimension are notoriously difficult 
to analyze, they are nonetheless highly relevant. In one dimension the two leg Hubbard ladder at half 
filling is the spin liquid epitome and as such ladder systems have attracted strong interest 
(for an early review see \cite{Dagotto}). Powerful analytic techniques such as bosonization and 
Bethe ansatz have been applied to single ladders with weak interactions (see \cite{ConTs} and 
references therein) and have led to comprehensive understanding of both doped and undoped ladders. 
In this letter we report the extension of these results to higher dimensions through the introduction of a 
small long range interladder tunneling in an ensemble of uncoupled half-filled Hubbard ladders. 
Increasing the tunneling amplitude leads to the formation of closed electron and hole Fermi pockets. 
The Luttinger Sum Rule (LSR) now takes on a novel form with the sign changes in the one electron Green's 
function appearing both as zeroes and infinities. This result has strong implications for the interpretation 
of ARPES results on underdoped cuprates. In the pseudogap phase the experiments interpret infinities as a 
set of disconnected Fermi arcs \cite{marnorm}, but do not (and cannot) observe the zeroes. Lastly we analyze possible 
instabilities of the carriers in the Fermi pockets using interactions derived from the low energy effective 
field theory for the ladders. 

The dynamics of the component half-filled ladders in our ensemble are governed at low energies by an
effective field theory.  As demonstrated in \cite{so8}, half-filled ladders with generically repulsive interactions 
experience under renormalization an enhancement in the symmetry of the bare Hubbard lattice
Hamiltonian.  With this enhancement, 
the effective field theory for the ladder is the SO(8) Gross-Neveu model, $H^{SO(8)}$ \cite{so8}.  
We couple the half-filled ladders together by long range single-particle tunneling.
The complete Hamiltonian is then
\bea
\sum_i H^{SO(8)}_i + \sum_{i,i' \atop l,l',n} 
\!\!t^\perp_{ii'll'}\!\!\int dx (c^\dagger_{nli\s}(x)c_{nl'i'\s}(x) + \mbox{h.c.}),
\eea
where $c^\dagger_{nli\sigma}$ creates an electron at the $n$-th site
on the $l$-th leg ($l=1,2$)
of the i-th ladder.
By making the hopping amplitude long range, we acquire a small parameter, as was done for a similar 
model of coupled Hubbard chains in \cite{EsTs}. 
We assume the following hierarchy of energy scales: $W ~{\rm (bandwidth)} \gg
\Delta ~{\rm (spectral~gap)} \gg t^\perp$.  The first inequality guarantees that the 
ladders can be described using the effective field theory.

The Gross-Neveu model is exactly solvable for all 
semi-simple symmetry groups and a great deal is known about its thermodynamics and correlation functions. 
In the SO(8) case the correlation functions were studied in \cite{KonLud,EsKon}. 
The spectrum of this model consists of three octets of particles of mass $\Delta$ and a 
multiplet of 29 excitons with mass $\sqrt 3\Delta$. 
Two octets consist of quasi-particles of different chirality transforming according to the 
two irreducible spinor representations of SO(8), while the third octet consists of vector particles. 
The latter include magnetic excitations as well as
the Cooperon (a particle with charge $\pm 2e$).
The 16 kink fields, carrying charge, spin, orbit, and parity indices, are direct descendants
of the original electron lattice operators on the ladders.
According to \cite{KonLud,EsKon},  
the corresponding single electron Green's function is given by     
\bea
G_{a}^{(0)}(\omega, k_x) = \left\{Z_a\frac{[\omega + \epsilon_a(k_x)]}
{\omega^2 - \epsilon_a^2(k_x) - \Delta^2} + G_{a,reg}\right\}, \label{G0}
\eea
where $a = \s,j$ with $\s$ spin and $j =\pm$ indexing the bonding/antibonding bands. 
$\epsilon_a(k_x)$ is the bare dispersion in the corresponding band. 
There are no off diagonal Green's function involving electrons from opposite Fermi points, 
a result of right and left moving quasi-particles belonging 
to different irreducible representations of the SO(8) group. 
As  was demonstrated in \cite{EsKon}, 
the incoherent part of the Green's function $G_{a,reg}$ yields a negligible 
contribution to the spectral weight.  Thus the quasi-particle weight satisfies $Z_a \approx 1$. 

\noindent{\bf RPA Analysis:} We will study the properties of our coupled ladders 
close to the Mott-Hubbard transition. Our approach follows closely the 
one developed in \cite{EsTs}. 
The interladder hopping is diagonal in the bonding/antibonding indices because of the bands' differing
Fermi wavevectors.
A Random Phase approximation (RPA) (diagrammatically pictured in Fig. 1)
yields the following expression for the full 2D Green's function:
\bea
G^{RPA}_a(k_x,{\bf k_\perp}) = \left\{(G_a^{(0)}(k_x))^{-1}  - t_a({\bf k}_{\perp})\right\}^{-1} .\label{G}
\eea
In our model the interchain tunneling amplitude has strong peaks at ${\bf k}_{\perp} =0, {\bf G}/2$,
where ${\bf G}$ is the inverse lattice vector in the direction perpendicular to the chains. 
(The peak at ${\bf G}/2$ follows from particle-hole symmetry, i.e. $t^\perp(k) = -t^\perp(k+{\bf G}/2)$.)
Near these points the following expansion is valid:
\begin{equation}
t_a({\bf k}_{\perp} + (1 \mp 1){\bf G}/4) = \mp t_{a0}[1 - ({\bf k}_{\perp})^2/\kappa_0^2 + ...],\label{t}
\end{equation}
where the dots stand for terms of higher order in $|{\bf k}_{\perp}|/\kappa_0$ and $\kappa_0 << G$ 
is the small parameter of the theory. 
We note that $t_{+0} >0$ (bonding) while $t_{-0} < 0$ (anti-bonding).

The quasi-particle spectrum is given by 
\bea
\omega - \epsilon_a(k_x) - \Delta^2(\omega + \epsilon_a(k_x))^{-1} - t_a
({\bf k}_{\perp}) = 0 .\label{secular}
\eea
At this point we note that the RPA Green's function (\ref{G}) together with (\ref{secular}) 
bears a remarkable resemblance to the 
single electron Green's function of underdoped cuprates conjectured in \cite{norman} 
on phenomenological grounds. 
In both cases the numerator of the self energy is modified.  In contrast, for a conventional superconductor
$t_a({\bf k_\perp})$ would be expected to modify $\epsilon_a(k_x)$.

\begin{figure}
\begin{center}
\epsfxsize=0.45\textwidth
\epsfbox{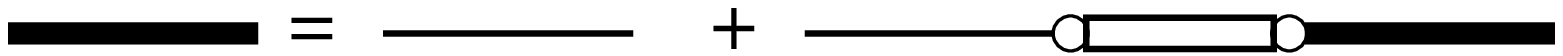}
\end{center}
\caption{The RPA equation for single particle Green's function (thick line), $G_a^{RPA}$.
The double line is the bare
hopping amplitude, $t_a$ while the thin line is the bare Green's function $G_0$.} 
\label{rpa}
\end{figure}

The dispersion relationships, $E_{a\pm}({\bf k})$, of the quasi-particles are given from Eq. \ref{secular} by
\bea
E_{a\pm}({\bf k}) = t_a({\bf k_\perp})/2 \pm \sqrt{(\epsilon_a(k_x) + t_a({\bf k_\perp})/2)^2 + \Delta^2}.
\eea
The FS are determined by solving $E_{a\pm}=0$.
Doing so to leading order for ${\bf k_\perp}$ near $0$ and ${\bf G}/2$ yields
\bea
[2(k_x - k_{Fa})v_F \mp t_{a0}]^2 + 2t_{a0}^2{\bf k_{\perp}}^2/\kappa_0^2 = t^2_{a0} - 4\Delta^2.
\eea
Gapless excitations, i.e. Fermi surfaces (FS), 
then only emerge when max$|t_a({\bf k_\perp})| > 2\Delta$.
For $\pm t_{a0} < 0$, the FS pockets are electron-like, while for $\pm t_{a0} > 0$, they are hole-like. 
Now $G^a_{RPA}= (\omega+\epsilon_a({\bf k}))/(\omega-E_{a+}({\bf k}))(\omega-E_{a-}({\bf k}))$.  
Near the FS of the pockets,
$\epsilon_a(k_x) \sim \pm\Delta$ and $E_{a+}({\bf k}) \sim \pm t_{a0}$.  The effective quasi-particle weight 
can then be read off to be
$ Z_{RPA} \sim 1/2 $. 
Thus RPA yields well defined quasi-particles.
\begin{figure}
\begin{center}
\epsfxsize=0.35\textwidth
\epsfbox{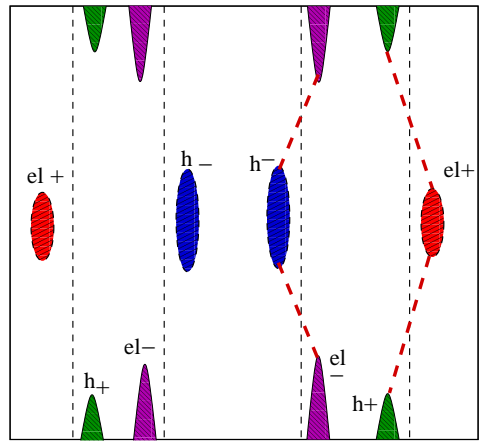}
\end{center}
\caption{Electron (red and magenta) and hole (green and blue) pockets. The difference in size 
between pockets formed from bonding and antibonding orbitals originates from possible difference between  
the corresponding tunneling amplitudes. The dashed lines represent the Luttinger surfaces $\epsilon_{\pm}(k)=0$.
The thick dashed red lines are loci of the gap minima, $\epsilon_a(k) = - t_a({\bf k})/2$.}
\label{FS}
\end{figure}

\noindent{\bf Luttinger Sum Rule:}
The Green's functions (\ref{G0}) and (\ref{G}) satisfy the Luttinger Sum Rule (LSR) in the form which, 
though being  well known theoretically, has had  limited applications. 
The LSR relates the electron density to the volume in momentum space in which $G(\omega=0,{\bf k}) >0$. 
This volume is bounded by the surface where $G(\omega=0,{\bf k})$ changes sign \cite{AGD}. 
The sign change can occur either at an infinity of $G(\omega=0,{\bf k})$ (Fermi surface) 
or a zero (Luttinger surface). 
The first possibility is denied for a Mott insulator.  For example, 
the Green's function (\ref{G0}) at $\omega = 0$ 
vanishes at $k_{Fa}$ where $\epsilon_a(k_{Fa}) = 0$, i.e. at momenta where the noninteracting Green's 
function had infinities. In this way the LSR is satisfied for this nonperturbative ladder ground state. 
In the presence of interladder tunneling, the Green's function (\ref{G}) continues to have zeroes at $k_{Fa}$ 
independent of the ${\bf k}_{\perp}$ component (see Fig. 2).  However, when Fermi surface 
pockets appear, the Green's 
function additionally changes sign through the newly formed infinities. 
Electron like pockets add and hole like pockets subtract from the total electron density, 
but the LSR remains valid. 
This example demonstrates, as does the case described in \cite{EsTs}, 
that in doped spin liquids it is generally 
necessary to determine both the Fermi and Luttinger surfaces in order to obtain the 
electron density from the LSR. 
It is important to point out that the Luttinger surface, determined by the zeroes of 
the Green's function, differs dramatically
from the surface of minimum gap (see Fig. 2). The latter is often used in ARPES 
experiments to extrapolate to an underlying 
Fermi surface. This however leads to difficulties in the pseudogap phase of 
underdoped cuprates where the enclosed 
area manifestly exceeds one electron per unit cell, inconsistent with hole doping (e.g. see \cite{shen}).

\begin{figure}
\begin{center}
\epsfxsize=0.35\textwidth
\epsfbox{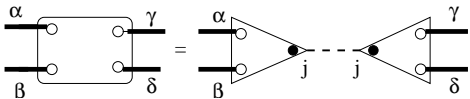}
\end{center}
\caption{Approximation of the four-quasi-particle interaction by the emission of an intermediate vector boson. } 
\label{43}
\end{figure}
 
\noindent{\bf Instabilities and Doping Dependence:} 
As was demonstrated in \cite{EsTs}, the RPA solution may become unstable at low temperatures. 
The instability is 
driven by the residual interactions between Fermi quasi-particles and collective modes of the 
spin liquid. This interaction 
may receive added strength from  nesting of the Fermi surfaces of particles and holes. 
To describe the instability one 
needs to move beyond RPA. We follow here the paper \cite{EsTs} and write down an effective 
action for quasi-particles 
interacting with collective excitations. The interaction comes from the diagram depicted 
on the l.h.s. of Fig.3. This four-point 
function can be approximated as shown on the r.h.s. of this same figure, leading to
the following effective action $(p= (\omega,{\bf k}))$: 
\bea
&&S = \frac{1}{2}\sum_{p,j}A_j(-p)[\omega^2 - (v_F k_x)^2 - \Delta^2]A_j(p) + \nonumber\\
&& \sum_{p,\alpha}\psi_{\pm}^{\bar\alpha}(-p)G^{-1}_{RPA}(p)\psi_{\pm}^{\alpha}(p) + \\\label{H}
&& \sum_{q,k} \Gamma(\frac{q}{\Delta},\frac{k}{\Delta})A_j(q)\psi_+^{\alpha}(k)
(C\gamma^j)_{\alpha\beta}\psi_-^{\beta}(-k -q)  ,\nonumber
\eea
where all fields  are real and $\bar\alpha$ is a charge conjugate of $\alpha$.  $A_j$ is the 
bosonic field transforming according to the vector representation of the SO(8) group, $\psi^{\alpha}_{\pm}$ 
are spinor fields of right and left chirality, and $\gamma^j$ are  gamma matrices of the SO(8) group. 
In principle,  there is an interaction within each particle multiplet, 
but we neglect it by accepting the approximation of Fig. 3.  
Such interactions lead to the creation of bound states 
with spectral gap $\sqrt 3\Delta$.  We, however, treat these as high energy processes. 
From general considerations, supported 
by the calculation that follows, we conclude that $\Gamma \sim \sqrt{v_F}\Delta$.  

In coupling the ladders together, the SO(8) symmetry is reduced to SO(6)
(provided $|t_{+0}|=t_{-0}|$.  The quasi-particles
that transformed in an 8 dim. representation under SO(8) now are arranged into 4 dim.
spinors.  These spinors are precisely the same that appear in the SO(5) theory of
superconductivity \cite{zhang}.  We thus expect the same phenomenology present in SO(5) models
to be present here.  At half-filling the coupling of the ladders will
lead to a spontaneous breaking of the SO(6) symmetry.  Possible ordered states
include superconductivity (SC), antiferromagnetism (AFM), and a staggered flux phase (SFP).  The physics
of explicit SO(6) breaking terms will be studied in later work.

Moving away from half-filling introduces a nonzero 
chemical potential $\mu$, and the SO(6) symmetry is reduced down to the SU(2)$\times$U(1). The chemical 
potential acts on the vector bosons as a ``magnetic field'' moving the Cooperon down in energy. 
At the same time it partially removes the nesting such that the electron and hole pockets become unequal in 
size and pockets of one type may even disappear. In this case SC becomes the leading instability. 

We are able to provide an estimate of the superconducting ordering temperature, $T_c$.
The RPA result remains valid at  $\mu < \Delta/2$ 
so that the Cooperon has not yet condensed and the ground state of the single ladder remains unchanged.  
(For $\Delta \approx 2\mu$ where Cooperons do condense, a more sophisticated 
approach similar to \cite{chubukov} is required.)
The dispersion relations
are modified to
\bea
&& E_{a}({\bf k}) \approx 
\frac{(k_x - p_0)^2}{2m_{\parallel}} + \frac{{\bf k}^2_{\perp}}{2m_{\perp}} - \epsilon_F - \mu ;\\
&& E_{a}({\bf k} + {\bf G}/2) 
\approx -\frac{(k_x - p_{{\bf G}/2})^2}{2m_{\parallel}} -\frac{{\bf k}^2_{\perp}}{2m_{\perp}} + \epsilon_F - \mu , \nonumber 
\eea
where $\epsilon_F = t_{a0}/4 - \Delta^2/t_{a0}$, 
$p_{(1\mp 1){\bf G}/4} = k_{Fa} \pm t_{a0}/2v_F$, $m_{\perp} = \kappa_0^2/t_{a0}$, and $m_{\parallel} = t_{a0}/2v_F^2$. 
In two dimensions the density of states on the Fermi surface 
is $\rho_F = \frac{\sqrt{m_{\parallel}m_{\perp}}}{\pi} = \frac{\kappa_0}{\sqrt{2}\pi v_F}$.  
The pairing susceptibility is
\bea
\chi^{-1} = [(\omega + 2\mu)^2 - (v_Fq_x)^2 - \Delta^2] + \Gamma^2\Pi(\omega,{\bf q}) \label{chi}.
\eea
Since the interaction decays at high energies at the scale $\Delta$, we can take it as the 
high energy cut-off 
in the polarization operator. At $\omega, q =0$ we have 
$\chi^{-1}(0,0) \approx -\Delta^2 + 4\mu^2 + (v_F\rho_F\Gamma^2/\Delta)\ln(\sqrt{(\epsilon_F + \mu)\Delta}/T)$ 
which determines the  mean field temperature of the transition to a superconductor with a stiffness 
determined by the dopant density: 
\bea
T_c \approx \sqrt{(\epsilon_F(t_{a0}) + \mu)\Delta}\exp\left[-\frac{1 - (2\mu/\Delta)^2}{\mbox{const}\times(\kappa_0 v_F/\Delta)}\right] \label{Tc}.
\eea
This expression is valid only when $t_{a0}$ exceeds $2\Delta$.
As $\mu$ increases so does $T_c$.  The $T_c$ of other possible instabilities
(AFM and SFP) is found approximately (ignoring the consequences of the destruction of a $(\pi,\pi)$ nesting)
by setting $\mu=0$ in Eqn. \ref{Tc}.  These instabilities are thus exponential disfavoured.

While the expression for $T_c$ is not valid if $\mu > \Delta/2$, we can still ask what
happens to the Fermi surface for temperatures above any putative $T_c$.
Upon increasing $\mu > \Delta/2$, the single particle gap {\it on the ladder} decreases but never
vanishes \cite{KonLud} and so 
the ladder Greens function retains its zeros.  It is these zeros that prevent the RPA
electron pockets from merging together (see Fig. 2).  We thus do not expect, within the validity of the model,
a transition to a large Fermi surface for some $\mu_c$.

To provide an estimate of $\Gamma$ entering the expression for $T_c$,
we calculate the three-point correlation function 
in the SO(8) 
Gross-Neveu model using the formfactor approach.   
To evaluate this correlator, we insert a resolution of the identity 
between fields, reducing the correlation function
to a sum over matrix elements.  Keeping only matrix elements 
involving single particle states we find upon Fourier transformation,
\begin{widetext}
\bea
&& \frac{1}{v_F^2}D_{\alpha\beta}^a(p_1,p_2) =  \frac{\la a,-p_1|A^a|0\ra}{E_1(E_1 + \omega_1)}\left[\frac{\la 0|\psi_+^{\alpha}|\bar{\alpha},-p_2\ra \la\bar{\alpha},-p_2|\psi_-^{\beta}|a,-p_1\ra}{E_2(\omega_2 - E_2)} + 
\frac{\la 0|\psi_-^{\beta}|\bar{\beta},-p_1-p_2\ra\la\bar{\beta},-p_1-p_2|\psi_+^{\alpha}|a,-p_1\ra}{E_{12}(\omega_1+\omega_2 + E_{12})}\right] + \nonumber\\
&& -\frac{\la 0|A^a|a,p_1\ra}{E_1(\omega_1-E_1)}\left[
- \frac{\la a,p_1|\psi_-^{\beta}|\bar{\beta},-p_2\ra \la\bar{\beta},-p_2|\psi_+^{\alpha}|0\ra}{E_2(\omega_2 + E_2)}  +
\frac{\la a,p_1|\psi_+^{\alpha}|\beta,p_1+p_2\ra\la\beta,p_1+p_2|\psi_-^{\beta}|0\ra}{E_{12})(\omega_1+\omega_2-E_{12})}\right] \label{3}\\
&& - \frac{\la 0|\psi_+^{\alpha}|\bar{\alpha},p_2\ra\la\bar{\alpha},p_2|A^a|\beta,p_1+p_2\ra\la \beta,p_1+p_2|\psi_-^{\beta}|0\ra}
{E_{12}E_2(\omega_2 - E_2)(\omega_1+\omega_2 - E_{12})} 
+ \frac{\la 0|\psi_-^{\beta}|\bar{\beta},-p_1-p_2\ra\la \bar{\beta},-p_1-p_2|A^a|\bar{\alpha},-p_2\ra\la\bar{\alpha},-p_2|\psi_+^{\alpha}|0\ra}
{E_{12}E_2(\omega_2+E_2)(\omega_1+\omega_2+E_{12})}.\nonumber
\eea
\end{widetext}
Each state is labeled by its isotopic index and momentum, $p$.  Momentum and energy are parameterized
in terms of rapidities, $\theta_i$ via $p_i = \Delta\sinh(\theta_i)/v_F$, $E_i=\Delta\cosh(\theta_i)$,
and $E_{12}=\sqrt{v_F^2(p_1+p_2)^2+\Delta^2}$.
The matrix elements of the Fermi operators are given by \cite{KonLud}:
\bea
&& \la 0|\psi_{\pm}^{\alpha}|\rho, \theta\ra = A\re^{\pm\ri\pi/4}C_{\alpha\rho}\re^{\pm\theta/2}; \\
&& \la\rho, \theta_1|\psi_{\pm}^{\beta}|a,\theta_2\ra = (C\gamma^aC)_{\alpha\rho}\re^{\pm(\theta_1+\theta_2)/4}g(\theta_1-\theta_2) ;\nonumber \\
&& g(\theta) = \frac{B}{1/2 - \cosh\theta} \times\nonumber\\
&&\exp\left\{\int_0^{\infty}\frac{\rd x\sin^2(x\theta/2\pi)}{x\sinh x\cosh(x/2)}\left[2\cosh(x/6) + \re^{-7x/6}\right]\right\},\nonumber
\eea
where $C$ is the charge conjugation matrix. 
$A$ and $B$ are related constants on the order of ${\sim \sqrt{\Delta/v_F}}$.  
For the Bose operators we have $\la 0|A^a|b,\theta\ra = \la b,\theta|A^a|0\ra = A_B \delta_{ab}, \la\rho|A^a|\eta\ra =0$,
where $A_B$ is ${\cal O} (v_F^{-1/2})$.
The vertex is then given in terms of the three point function via
$\Gamma^a_{\alpha\beta} = 2\pi G_a^{-1}(p_1)G_\alpha^{-1}(p_2)G_\beta^{-1}(-p_1-p_2)D^a_{\alpha\beta}(p_1,p_2)$ where 
\bea
&&D_{\alpha\beta}^a(p_1,p_2) = 
\frac{A_B A (C\gamma^a)_{\alpha\beta}\re^{-\theta_{12}/4}g(\theta_{12})}{E_1E_2}\times\\&&
\hskip -.2in \left[\frac{\re^{-\ri\pi/4}}{(\omega_1 \!-\! E_1)(\omega_2 \!+\! E_2)} \!-\! \frac{\re^{\ri\pi/4}}{(\omega_1 \!+\! E_1)(\omega_2 \!-\! E_2)}
\right]\bigg|_{{{\frac{p_1}{\Delta}}=\frac{\sinh(\theta_1)}{v_F}}\atop{-\frac{p_2}{\Delta}=\frac{\sinh(\theta_2)}{v_F}}} \nonumber  \\
&& - ((p_1,p_2,\omega_1,\omega_2) \rightarrow (-p_1,p_1+p_2,-\omega_1,-\omega_1-\omega_2)), \nonumber
\eea
and the $G's$ are the corresponding non-interacting propagators.
We also note that $g(x) \sim -B\re^{-|x|/4}$ for $|x| >> 1$. 
As we see, the  vertex is a smooth function of momenta and frequencies changing 
with a characteristic scale $\Delta$ as written in (\ref{H}). 
This derivation justifies Eqs. (\ref{Tc}).

In conclusion, we have constructed a toy model of a doped spin liquid.  This model possesses a number of
interesting features.  Vis-a-vis ARPES measurements, it offers an alternative framework in which to understand the
observed arcs in underdoped cuprates \cite{shen,marnorm}: such arcs may be Fermi pockets unresolved
by ARPES due to disorder and limited experimental accuracy.  It further suggests using an observed line of minimal
gap will lead to overestimate of the number of electrons present in a band.  Beyond implications for ARPES, 
the model (at half-filling) possesses an 
SO(6) symmetry and so encompasses the same phenomenology as SO(5) models of superconductivity including
a $\pi$-resonance at energy $2\mu$.  Away from half-filling 
superconductivity is preferred and the model is under sufficient control to 
provide an estimate for the superconducting $T_c$.  Finally we point out above $T_c$ the model predicts
the existence of a low lying Cooperon excitation.  This excitation's existence and its concomitant 
near coherent
propagation provides a means to understand
the giant proximity effect seen in LSCO/LCO/LSCO thin films
where the Josephson current was measured through LCO in its normal state \cite{bozovic}.
 
RMK and AMT  acknowledge the support
from US DOE under contract number DE-AC02 -98 CH 10886. TMR 
acknowledges hospitality from the Institute for Strongly
Correlated and Complex Systems at BNL. 

\end{document}